\documentclass[pre,twocolumn]{revtex4}
\usepackage{graphicx,t1enc}
\usepackage{amssymb,amsmath}
\usepackage[utf8]{inputenc}
\usepackage{verbatim}
\begin{document}

\title{A random walk model to study the cycles emerging from the exploration-exploitation trade-off}

\author{Laila D. Kazimierski}
\author{Guillermo Abramson}
\author{Marcelo N. Kuperman}

\affiliation{Consejo Nacional de Investigaciones Cient\'{\i}ficas y T\'ecnicas \\
Centro At\'omico Bariloche (CNEA) and Instituto Balseiro, R8400AGP Bariloche, Argentina.}

\begin{abstract}
We present a model for a random walk with memory, phenomenologically inspired in a
biological system. The walker has the capacity to remember the time of the last visit to each site
and the step taken from there. This memory affects the behavior of the walker each time it reaches
an already visited site modulating the probability of repeating previous moves. This probability
increases with the time elapsed from the last visit. A biological analog of the walker is a
frugivore, with the lattice sites representing plants. The memory effect can be associated with the
time needed by plants to recover its fruit load. We propose two different strategies, conservative
and explorative, as well as intermediate cases, leading to non intuitive interesting results, such
as the emergence of
cycles.
\end{abstract}

\date{\today}

\maketitle

\section{Introduction}
The movement of animals in search for food, refugia or other resources is nowadays the subject of
active research trying to unveil the mechanisms that give rise to a wide family of related complex
patterns. In particular, physicists find in these a fruitful field to explore reaction-diffusion
mechanisms~\cite{kuperman,abramson2013}, to apply the formalism of stochastic differential
equations~\cite{okubo02,mikhailov06,schat96} and to perform simulations  based on random
walks~\cite{viswanathan11,viswanathan96,giuggioli09,borger08}. 

One of the key aspects of this phenomenon is the fed back interaction between the individual and the
environment \cite{turc98}. These interactions may involve intra and inter specific competition
that, together with previous experience \cite{nath08,mor10} and the search for resources, drive
the displacement of the individuals. In particular, when animals move around in order to collect
food from patches of renewable resources, their trajectories depend strongly on the spatial
arrangement of such patches \cite{ohashi07}. This observation has motivated a large
collection of studies focused on finding optimal search strategies under different assumptions of
animal perception and memory \cite{bartumeus02,fronhofer13}. A related open question is that of the origin of home ranges, a concept introduced in \cite{burt} 
to characterize the  spatial extent of  the displacements of an animal during its daily activities.
Many species perform bounded explorations around their refugia, even though the available space and
resources extend far beyond. There are several 
hypotheses that try to explain this phenomenon, which could be only an emergent behavior 
associated to very simple causes \cite{abramson2014}. The review by B\"orger et al. 
\cite{borger08} is an exhaustive compilation of the state of the art. 
There, the authors  point out that movement models not always lead to the formation of stationary 
home ranges. 
Still, home ranges arise, for example, in biased diffusion \cite{okubo02}, in self-attracting walks 
\cite{tan} and in models with memory \cite{schu}. Nevertheless, the quest to unveil and characterize 
the underlying weave of causes and effects
behind  the  emergent patterns is not over. How do these emerge as the result of the interaction between the
behavior of an organism and the spatial structure of the environment?

In this context, the venerable symmetrical random walk has been the subject of many studies, with a
large collection of applications and characterizations that  include aspects
beyond the simple walker capable of only uncorrelated short-range steps. Just to focus on what we
want to present here, let us restrict the examples to random walks on discrete lattices where the
walker can gather information to build up a history. One such case is the self avoiding walk (SAW),
where the walker builds up its trajectory by avoiding to step onto an already visited site
\cite{flor,dege}. A characteristic result corresponds to the walker running into a site
with all its neighboring sites  already visited and being blocked. The converse case occurs when the walker
prefers sites visited earlier.

Previous works have shown that introducing long-range correlations into a random walk may lead to non
trivial effects translated into drastic changes in the asymptotic behavior. The usual diffusive
dynamics can evolve into  sub-diffusive, super-diffusive or persistent. Such random walks with
long-range memory have been extensively studied in recent years
\cite{hod,schu,trimp99,trimp01,kesh,para, silva,cres}.

In \cite{sapo,orde1,orde2} a behavior that can be interpreted as memory has been explored. These works analyze a self attracting  walk where
the walker jumps to the nearest neighbor according to a probability that increases
when the site has already been visited. A generalization that includes an enhancement of this memory with the frequency of
visits, but also with a degradation with time, was proposed in \cite{tan}.

In this work we propose a random walk with a specific memory that induces local correlations at long
times. The rationale for this model is to mimic the movement of a foraging animal, e.g. a frugivore,
going from one plant to another in order to feed. We show that the emergence of looped walks,
that can be associated with home ranges, can be promoted by very rudimentary capacities of
the individual together with a natural dynamics of the environment.

\section{The model}

For a  forager the proximity of a plant
is not enough to make it attractive for a future visit: the plant  must also have a visible and 
interesting load of fruit. 
Moreover, when visiting a plant the animal usually takes only part of the available fruit and moves 
on. After this, the plant needs some time to recover its fruit
load. Such a model was analyzed in \cite{abramson2014}. We attempt here a further simplification, 
coding the complex interaction of memory, consumption and relaxation in the probabilities defining 
the random walk from each site of the lattice. 

As a first simplification, consider that the animal eats all the available ripe
fruit in the visited plant and leaves. Let us say that a walker moving in such a substrate has a memory, allowing it to
remember the time of visit to every site and the step taken from there. When revisiting a fruitful
plant the animal will consider it a success and repeat the step taken from there, ``remembering''
its previous visit. When returning to a plant before its recovery the walker takes a random step.
This unlimited memory is not necessary associated to an extraordinary skill of the forager. It  could be 
stored in the environment as the state of each plant, which proximity and fruit load can trigger 
on the forager the inclination to choose a specific direction. Thus, the memory of having 
visited a site once, needs not to be stored on the animal but recorded on the topology of the 
environment (as is the case in \cite{abramson2014}). Also, we can anticipate here that when a home range emerges the walker effectively uses a bounded amount of memory.

Besides this, imagine two possible strategies for the \emph{update} of the memory, the details of
which will be given below. A \emph{conservative} walker will keep in memory the time in which the
visit to that site was successful and the step taken on that occasion. An \emph{exploring} walker,
instead, will update the memory of the visit to the current time and the step to the
randomly chosen one. Between these two strategies there might be intermediate ones, all of which 
will be explored below.

Now, with the motivation just exposed, let us define a random walk that modifies the probabilities of steps from each site according to the time since the last visit and a parameter defining the strategy. The rules of the walk can be summarized as follows:
\begin{itemize}
\item When visiting a new site, take a random step in either of the four directions. Store in memory the time of visit $t_v$ and the step.
\item When returning at time $t$ to a site previously visited at time $t_v$:
\begin{itemize}
\item With probability $p_r(t-t_v)$ repeat the step stored in memory. Update the visit
time stored in memory. 
\item[Or:]
\item With probability $1-p_r(t-t_v)$ take a random step and:
\begin{itemize}
\item With probability $\rho$, update in memory the time of visit and the step taken.
\item[Or:]
\item With probability $1-\rho$, keep the memory unmodified.
\end{itemize}
\end{itemize}
\end{itemize}
The probability distribution used to repeat the step taken in the previous visit is used to model
the replenishment of the fruit mentioned above. It can be simply a Heaviside step function
$p_r(t-t_v)=\theta(t-t_v-\tau)$, where $\tau$ is a parameter representing the recovery time of the
plants. It is equivalent to the memory of the \textit{elephant walk}~\cite{schu}, but used
in a different way. Contrary to the usual memory that makes the probability of revisiting a
site fade out with time, here we are considering a probability of revisiting a site that increases
with time. In such a case the walker will always repeat its step when returning after $\tau$ steps,
and always take a random step when returning earlier. This strict condition can be relaxed by
modeling $p_r$ with a smooth step function. In the results shown below only the Heaviside step 
distribution will be used, since, as we will show later,  no significant differences where found 
when using a smooth distribution. In such a case, the walks are characterized by two parameters, 
$\tau$ and $\rho$. 

Our results show the emergence of closed circuits in non trivial ways. To characterize the behavior
of these we analyze both the duration of the transient elapsed until the walker enters the closed circuit,
as well as  the length of such cycles.

The emergence of such circuits is reflected in the fact that during the initial stages the mean
square displacement exhibits a diffusive behavior whereas for longer times it reaches a plateau. 
Such a behavior has been already reported in previous works \cite{trimp99, trimp01} 
where due to a fed back coupling between a particle and its environment, it gains experiences
with modified surroundings, resulting in a bounded walk.

\section{Results}
The results presented below correspond to mean values taken over $10^3$-$10^4$ realizations, on
a 2 dimensional lattice, large enough to avoid that the walker reaches the borders. The simulations 
were done for
$10^5$ and $10^6$  time steps, showing no significant dependence between them.

\begin{figure}[t]
\centering
\includegraphics[width=\columnwidth, clip=true]{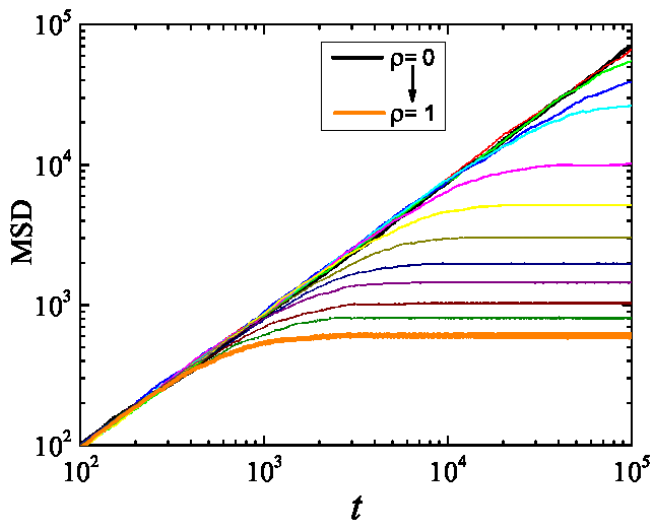}
\caption{Mean square displacement vs. time for probability $\rho=0$ (black), $\rho=1$
(orange) and intermediate values, and corresponding to recovery time of
the plants $\tau=20$. Simulations performed in a square lattice of $5000\times 5000$
sites, $10^5$ time steps and $10^3$ realizations. (Color on line.)}
\label{figure:MSD_tau20}
\end{figure}

One of  most revealing features of any sort of walk, be this random, self avoiding, self
attracting, etc. is its mean square displacement (MSD). The behavior of the MSD in the present model shows rather interesting features. Figure
\ref{figure:MSD_tau20} displays the MSD as a function of time for a range of values of
$\rho$, from 0 to 1, and for
$\tau=20$. Recalling that $\rho$ is the probability that the walker updates the information,
stored in its memory, regarding the time of visit to a site and the step taken from there, we
associate $\rho=1$ with
the \emph{exploring} behavior and $\rho=0$ with the \emph{conservative} one. We observe that for
$\rho=0$ the behavior is
clearly diffusive, while for $\rho=1$ the MSD reaches a plateau indicating that the walker remains
trapped in a bounded region. Contrary
to the intuitive guess, this shows that it is the exploring behavior the one which allows the
walker to find closed circuits more easily, while the conservative behavior leads to a
diffusive walk. Intermediate values of $\rho$ generate intermediate behaviors.
We have analyzed the model for values of $\tau$ ranging from 5 to 150, finding analogous results for all of them.

These results rise several questions about the dependence of the the emergence of cycles
on each parameter. Even though all two-dimensional walks (including the case $\rho=0$) eventually return to a site in a
condition that allows the settling of a cycle, the time necessary to fulfill this condition can vary
greatly. As a result of this, after a fixed number of steps only
a fraction of the walkers are able to do so. In the following we proceed to characterize the
statistical behavior of these walkers by measuring several relevant quantities.

\begin{figure}[t]
\centering
\includegraphics[width=\columnwidth, clip=true]{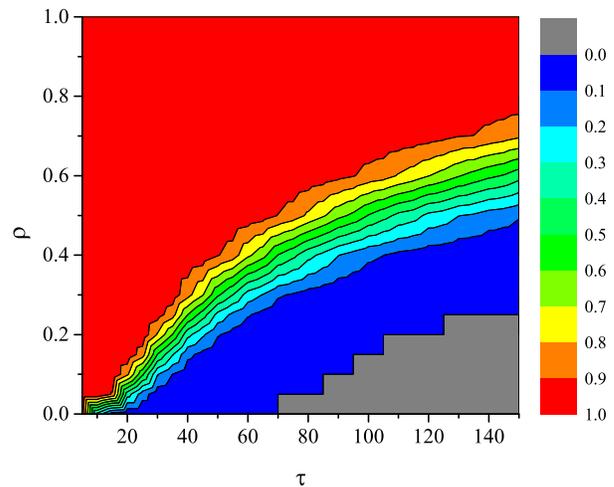}
\caption{Contour plot of the fraction of realizations that end in a cycle, as a function of 
parameters
$\rho$ and $\tau$.
Simulations performed in a square lattice of $5000\times 5000$ sites, $10^5$ time steps
and $10^4$ realizations. The gray region corresponds to realizations that do not end in cycles due 
to finite observation time.}
\label{figure:cant_ciclos}
\end{figure}

Figure~\ref{figure:cant_ciclos} shows a contour diagram representing the fraction of realizations that
end in a cycle, as a function of the parameters $\rho$ and $\tau$. We observe that this fraction increases both for decreasing $\tau$ as well as for increasing $\rho$.
Consistently, mapping this situation to the biological scenario, when plants take too long to
recover (large $\tau$), or when the foragers are not exploring enough (too small $\rho$), there is no formation of home ranges.

\begin{figure}[htp]
\centering
\includegraphics[width=\columnwidth, clip=true]{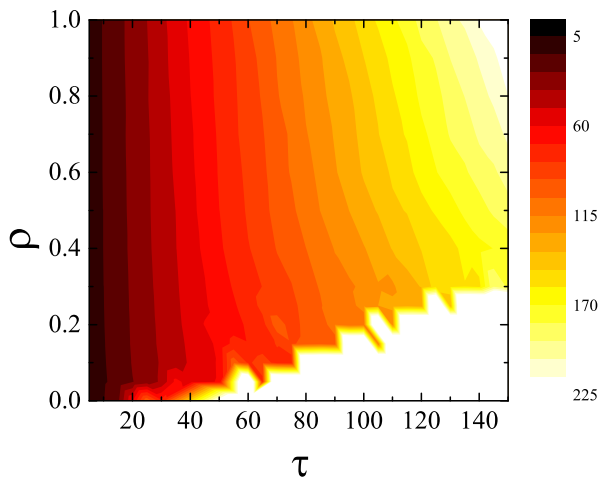}
\caption{Contour plot of the mean cycle length as a function of $\rho$ and $\tau$.
Simulations performed in a square lattice of $5000\times 5000$ sites, $10^5$ time steps
and $10^4$ realizations.}
\label{figure:prom_ciclos}
\end{figure}

Another informative aspect of the walks that needs characterization is the length of the cycles. 
The concept of a home range is always associated to the measurement of the amount of space 
utilized. Sometimes  it is measured through the utilization distribution \cite{ford79}, that 
represents the probability of finding an animal in a defined area within its home range.
In this case, once the cycle is established, the animal will visit each site within the cycle 
only once at each turn, so the utilization distribution will be uniformly distributed among the 
sites within the cycle. Still we can have an estimation of the amount of used spaced by 
measuring the longitude of the cycle.
A priori we know that $\tau$ is the greatest lower bound (infimum) for the average cycle length.
This average is shown in Fig.~\ref{figure:prom_ciclos}. We can conclude that the mean length of the
cycles is very close to this bound for all parameters sets, showing a very weak dependence on $\rho$
for the largest values of $\tau$, undoubtedly due to the undersampling arising from the finite
simulation runs. Observe, nevertheless, the
wedge shaped region of very conservative walkers that never find a cycle, which grows with the
recovering parameter $\tau$.

\begin{figure}[htp]
\centering
\includegraphics[width=\columnwidth, clip=true]{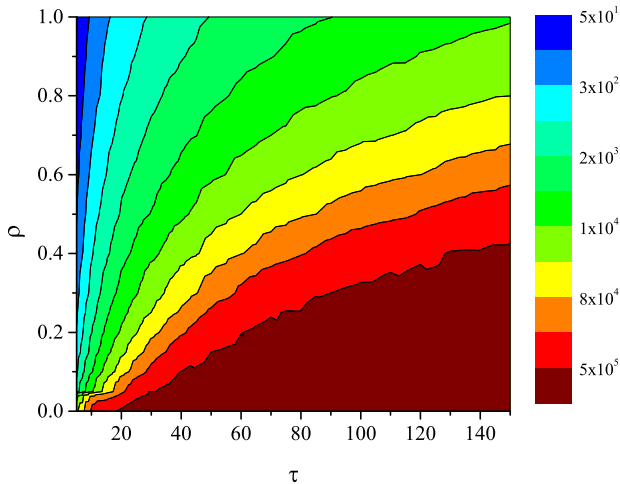}
\caption{Contour plot of the mean transient length as a function of parameters
$\rho$ and $\tau$. The color scale is logarithmic. Simulations performed in a square lattice of 
$10^4\times10^4$ sites, $10^6$ time steps.}
\label{figure:prom_transitorio}
\end{figure}

\begin{figure}[htp]
\centering
\includegraphics[width=\columnwidth, clip=true]{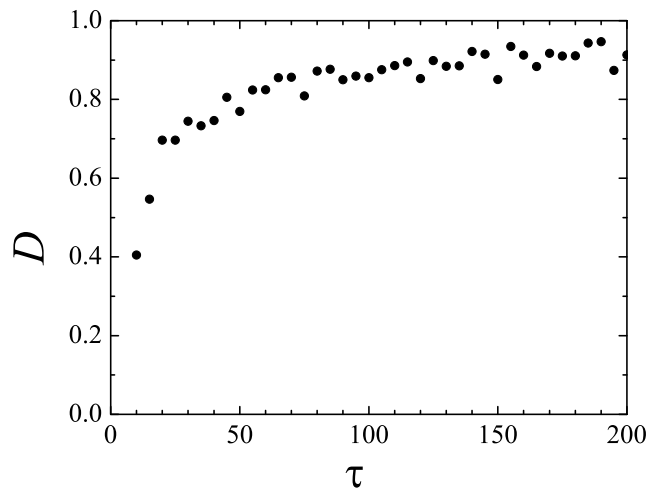}
\caption{Diffusion coefficient of the $\rho=0$ case (slope of the average MSD curves for each value 
of $\tau$). 
Forty uniformly distributed values of $\tau$ were considered
between 5 and 200. }
\label{figure:d_vs_tau}
\end{figure}

Let us now focus on the extreme cases of $\rho=0$ and $\rho=1$. When $\rho=0$ we found that the
behavior is diffusive for all values of $\tau$, so that
$\langle x^2\rangle =D(\tau)\,t$. As shown in Fig.~\ref{figure:d_vs_tau}, $D(\tau)$ depends on
$\tau$ approaching 1 from below as $\tau$ increases. On the other hand, perfect explorers---those
with $\rho=1$---always find a cycle. We have found that the average length of the transient depends 
quadratically on $\tau$.

The transient regime is longer as the value of $\tau$ is larger, i.e.,
for short recovery times, the walker finds  a cycle easier (and faster).
If $\tau$ is very large, it may happen that the walker returns successive times to the same site
earlier than $\tau$, and randomly choose the next step, losing the possibility of repeating
the last steps and thus entering a cycle.

Observe that the exploring walker is the one that continuously updates the stored information.
An intuitive guess of the resulting dynamics, analyzed in terms  of the
intensity of exploring activity of the individual, may lead us to think that such
walker would have a higher difficulty in
establishing
a walking pattern and finding a closed circuit. Also, for those who maintain the stored
information (the conservative walkers), finding an optimal closed circuit would be a relatively 
simple task.
However, our results show that this intuition is wrong.

Relevant insight on the mechanisms that give rise to the observed behavior of the forager walk can
be obtained from well known results of conventional random walks. A random walk in one and two
dimensions is recurrent, i.e. the probability that the  walker eventually returns to the
starting site is 1. (In  higher dimensions, the random walk is transient, the former probability
being less than 1 \cite{green}.)
So, in principle, for any value of $\tau$ and $\rho=0$ the forager walk eventually ends up in a
cycle. However, this asymptotic behavior of the system may not be the most relevant one in many
contexts. In the biological scenario, for example, one would be interested in the possibility of
finding cycles in
relatively short times.

Our results can be explained by considering the so called \textit{Pólya problem}
or first return time.  The probability that a simple random walk in one dimension returns for the
first time to a given site after $2n$ steps is
\begin{equation}
\binom{2n}{n}\frac{1}{(2n-1)2^{2n}}.
\label{first}
\end{equation}
In two dimensions the probability that a simple random walk returns to a given site  after
$2n$ steps is the square of the previous probability  \cite{green} as a
simple random in two dimensions can be projected into two independent
one-dimensional walks on the $x$ and $y$ axes. The probability given by Eq.~(\ref{first})
asymptotically decays as $n^{-3/2}$, indicating that returning to the initial site is increasingly
improbable with the elapsed time.
The forager walk can be interpreted in  the following way. Until the moment that the walker gets
trapped in a cycle, it performs a random walk. Afterwards, the behavior is deterministic. That very
moment corresponds to the first time a cycle is completed, so it is a return to the initial step of
the cycle after $\tau_c\ge \tau$ time steps, where $\tau_c$ is the period of the cycle of an
individual realization for a given choice of $\tau$. Let us assume that the transient walk executed
up to this first return can be used to estimate a probability analogous to Eq. (\ref{first}). We can
do this from the length of the transient and the fraction of realizations that successfully ended in
a cycle.
The transient can be thought of as consisting of successive realizations of walks of length 
$\tau_c$ 
that were \emph{not} successful in returning to the starting point. We have verified this algebraic 
dependence.

The immediate question about the validity of the present results for higher dimensions can be 
answered by invoking the recurrence theorem presented by G. Pólya in  1921 \cite{poly21}, where he. 
shows that a random walk is recurrent in 1 and 2-dimensional lattices, and that it is 
transient for lattices with more than 2 dimensions. 
The emergence of home ranges as presented in this work is strongly dependent of the 
probability of eventual returns to already visited places. Thus, for dimensions higher than 2 the 
expected cycle lengths will be longer and their very existence less probable, as can be deduced from 
the calculated probabilities of returns to the origin in these cases \cite{mont56}.

Besides, the fact that increasing $\rho$ produces an increase in the
probability of finding a cycle can be understood in the following way. The probability
of returning to a given site decreases as the walker moves away. When $\rho$ is small the walker can
move increasingly
farther away from the stored site, making it rather difficult to return to it and enter a cycle.
When $\rho$ is high the foraging walker constantly updates its memory, in a way that it is always
relatively close to the most recently stored site. This increases the probability of returning to it
and triggering a cycle.

\begin{figure}[htp]
\centering
\includegraphics[width=\columnwidth, clip=true]{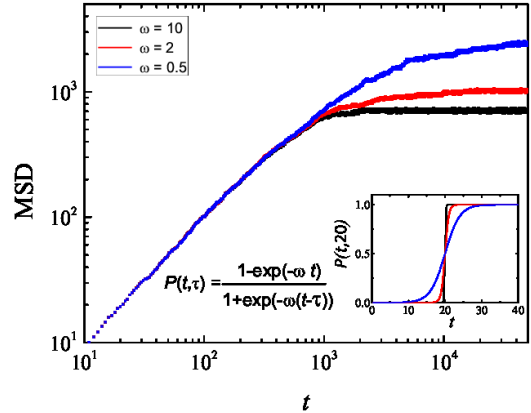}
\caption{Mean square displacement vs. time considering a smooth function for the probability 
to repeat the step taken in the previous visit.
$\omega=10$ (black), $\omega=2$ (red), $\omega=0.5$ (blue) and $\tau=20$ (color on line). 
Simulations performed in a square lattice of $5000\times 5000$
sites, $10^5$ time steps and $10^3$ realizations. 
The inset shows the functional expression and shape of the probability distribution.}
\label{figure:nostep}
\end{figure}

For completeness, we  include a plot showing results based on the use of a smoother distribution. 
The smooth step depends on two parameters, $\tau$ and $w$. The limit $w\to\infty$ tends to a 
Heaviside step function at $t=\tau$. Figure \ref{figure:nostep} displays the behavior of the walk 
for three values of $\omega$ ($10$, 2 and $0.5$), exemplifying the typical behaviors for a fixed 
value of $\tau=20$. The MSD's are averages over 1000 realizations. The black curves correspond to 
$\omega=10$, which is very similar to a step, and gives an MSD almost identical to the one shown in 
Fig. 1, with $\rho=1$ (orange curve). While smoother curves tend to plateaus at higher values
no qualitatively differences  are observed in the behavior.

\section{Conclusions}

One of the important aspect related to animal movement is the  effect that spatial heterogeneities 
have on the observed patterns. When the spatial heterogeneity is manifested 
through the distribution of resources, the  link between resource dynamics and random walk models 
might be the key to answer many of open questions about the emergence of home ranges. Another route 
to explore this problem is by accounting for learning abilities and spatial memory  \cite{stamp99}.

The formation of a home range has previously been investigated with models in which a single 
individual displays both an avoidance response to recently visited sites  and an attractive 
response toward places that have been visited sometime in the past \cite{fronhofer13,moor09}. 
An animal searching for food would choose its movements based not only on its internal state and 
the instantaneous perception of the environment, but also
on acquired  knowledge and experience. Animals use their memory to infer the current state of 
areas not previously visited. This memory is build up by collecting information remembered from 
previous visits to neighboring locations \cite{faw14}. 

Although the emergence of home ranges is crucial in understanding the patterns arising from animal movement, there are few mechanistic models that reproduce this phenomenon.
Traditional random walks, widely used to describe animal movement, show a diffusive behavior 
far from displaying a bounded home range. However, the addition of memory capacity 
has proven to predict bounded walks \cite{schu,trimp99,trimp01}.
Home ranges also arise in biased diffusion \cite{okubo02} and in self-attracting walks 
\cite{tan}. The interest aspect of the results presented here is that they not only reveal the 
non trivial behavior of the so called \textit{frugivore walk} but also contribute to a deeper 
understanding of the causes underlying the constitution of home ranges as an emergent phenomenon, 
among which we highlight the foraging strategy. By considering a minimal model we have shown that a 
walker with rudimentary learning abilities, together with the feedback from a dynamic substrate, 
give rise to an optimal foraging activity in terms of the usage 
of the spatial resource. Indeed, neither a foraging strategy based just on diffusion (a random walk 
without memory), nor a walk strongly determined by memory (like our conservative walker), are 
optimal. A better strategy is one that combines the use of memory with an exploratory behavior, 
such as our \emph{explorative} walker. 

There is evidence supporting that precisely this combined strategy may be the one favored by 
evolutionary mechanisms \cite{gaf81,eli07}.
Foraging activity must balance between exploration and exploitation: on the one hand, exploring the 
environment
is crucial to find and learn about the distributed resources; on the other hand, exploitation of 
known resources is
energetically optimal.
Indeed, this trade-off is a central thesis in current studies of foraging ecology, as it is 
apparent in the thorough work by W. Bell \cite{bell1991}, in Lévy flight models \cite{viswanathan11} 
and others.
The simple mechanism analyzed here contributes with theoretical support to these ideas. We have shown 
that the balance between exploration and exploitation not only provides an optimal use of 
resources. 
It may also be responsible for the emergence of a home range. The balance between exploration and 
exploitation appears as the road to successful foraging.

\begin{acknowledgments}
This work was supported by grants from ANPCyT (PICT-2011-0790), U. N. de Cuyo (06/C410) and CONICET
(PIP 112-20110100310)
\end{acknowledgments}

\end{document}